\documentstyle[preprint,prl,aps]{revtex}
\begin{document}
\draft

\title{Random Bonds and Topological Stability in Gapped Quantum Spin Chains}

\author{R. A. Hyman$^1$, Kun Yang$^2$, R. N. Bhatt$^2$ and S. M. Girvin$^1$}
\address{$^1$Physics Department, Indiana University, Bloomington, IN 47405}
\address{$^2$Department of Electrical Engineering, Princeton University,
Princeton, NJ 08544}
\date{\today}

\maketitle

\begin{abstract}
We study the effects of random bonds
on spin chains that have an excitation
gap in the absence of randomness.  The dimerized
spin-1/2 chain is
our principal example. Using an asymptotically exact real space
decimation renormalization group procedure, we find that dimerization
is a relevant perturbation at the random singlet fixed point.
For weak dimerization, the dimerized chain
is in a Griffiths phase with short
range spin-spin correlations and a
divergent susceptibility.
The string topological order, however, is not destroyed by bond
randomness
and dimerization is stabilized by the confinement of topological defects.
We conjecture
that
random
integer spin chains in the
Haldane phase exhibit similar thermodynamic and topological properties.
\end{abstract}

\pacs{Pacs: 75.10.J, 75.30.H, 75.50.E}
Extensive theoretical work on random quantum
magnetic systems has been carried out since
the
late 1970's
\cite{mdh,bl,fisher0,df,fisher1,fisher2}.
Systems that behave critically in the
absence of randomness are unstable against
weak randomness and flow to
the random singlet (RS)
phase\cite{mdh,fisher1}. In the RS phase,
spins far apart in
space form weakly bound
singlet pairs in a more or less random manner.  This phase is
also referred to as the valence bond glass (VBG) phase\cite{bl}.
The low temperature thermodynamic properties
of these systems are
dominated by the
weakly bound pairs and are universal\cite{mdh,bl,fisher1}.
For instance, the susceptibility of the undimerized random
antiferromagnetic Heisenberg or $XXZ$
spin-1/2 chain diverges as $[T\log^2T]^{-1}$ at low $T$, independent of the
details of the randomness\cite{fisher1}.
Universal power law behavior has also been found in disorder averaged
spin-spin correlation
functions\cite{fisher1}.
Experiments,
however, seem to find power law divergent
susceptibilities with nonuniversal exponents\cite{experiments}.
It is
of interest to study if there exist relevant perturbations
at the RS fixed point that
drive the system
towards a state exhibiting the nonuniversal behavior found
experimentally.

A related issue is the effect of randomness on spin chains that have
an excitation gap in the absence of randomness\cite{sorensen}.
The most prominent examples of such chains are
integer spin chains in the Haldane phase\cite{spin1}.
Other examples include
dimerized spin-1/2 chains\cite{singh} and spin chains with
spontaneous dimerization\cite{next-near,nomura}.
All of these systems have topolologically ordered\cite{rn} ground states.
One might think that strong enough randomness will inevitably
destroy the topological order of the ground state.
However Haldane has suggested\cite{duncan}
that there exists a class of random perturbations for which the
the topological order in the
ground state of integer spin chains is stable
regardless of the strength of the perturbations.  We will show that
the analog of this prediction for random bond dimerized
spin-one-half chains is, in fact, correct.

An explicit example that provides strong support to
Haldane's conjecture (sic!)
can be found in the random version of the AKLT model:\cite{aklt}
\begin{equation}
H=\sum_i{J_i\left[{\bf S_{i}}\cdot{\bf S_{i+1}}
+{1\over 3}({\bf S_{i}}\cdot{\bf S_{i+1}})^2\right]},
\end{equation}
where $J_i>0$. The exact
ground state of this random model is identical to that
of the pure model,\cite{aklt} i.e., a valence bond solid.
Its excitation
spectrum and
thermodynamic properties will certainly depend on the distribution
of $J_i$, yet the perfectly topologically ordered ground state
is completely unaffected by randomness.

In the absence of disorder, spin-one chains in the Haldane phase and dimerized
spin-one-half chains exhibit similar physical properties\cite{hida}.
They both have a nondegenerate
ground state with an excitation gap, and more importantly, they both
have string-topological order\cite{hida}.
Hida has shown that they can
be continuously connected to each other without closing the gap
or removing the topological
order; i.e., they are in the same phase.\cite{hida}
It is natural to expect
that
they also behave similarly in the presence of randomness.

In this paper we study the random bond dimerized spin-1/2 chain
in detail.
Using the asymptotically exact
real space decimation renormalization group introduced
by Ma, Dasgupta and Hu\cite{mdh} and extended by Fisher\cite{fisher1},
we find that enforced dimerization is
a relevant operator at the RS fixed point
that drives the system to a random dimer (RD) phase.
The low temperature
thermodynamic properties of the RD phase are
{\em nonuniversal}.
For weak dimerization, the spectrum of the RD
phase is gapless and the susceptibility diverges as $\chi\sim T^{-1+\alpha}$
with $\alpha>0$ and dependent
on the bond distribution,
(similar to the behavior found experimentally and in qualitative agreement
with the RS thermodynamics),
but the averaged
spin-spin correlation
function remains short ranged.
Thus for weak dimerization the RD phase is an example of a Griffiths phase.
More importantly, we find that the string-topological order is {\em not}
destroyed by random bonds.
We conjecture that these results also
apply to random bond integer spin chains
in the Haldane gapped phase.
Comparison will also be made with spontaneously
dimerized spin-chains
which behave very differently
upon introducing disorder.

Consider the model Hamiltonian
\begin{equation}
H=\sum_i{J_i\left[S_i^xS_{i+1}^x+S_i^yS_{i+1}^y+\Delta S_i^zS_{i+1}^z\right],}
\label{model}
\end{equation}
where $S_i^\alpha$ are
spin-one-half operators, $J_i$ are (random) positive
coupling constants, and $0\le \Delta\le 1$.
Here we will concentrate on the cases $\Delta=0$ ($XX$ chain\cite{notexx}) and
$\Delta=1$ (Heisenberg chain), since it has been shown that, for the
case of random bonds,  the Ising coupling
is irrelevant when $\Delta<1$ and that the system flows to the random
$XX$ chain in
the low energy limit\cite{fisher1}.
We assume that the distribution of the couplings, $J_i$,
depends on whether $i$ is even or odd.
We write the distribution functions for even and odd $J$'s as $P_e(J, J_0)$
and $P_o(J, J_0)$ respectively. Here $J_0$ is the cutoff in the distribution
function corresponding to the strongest bond in the system.
As Fisher\cite{fisher1} has shown, in the absence of dimerization, i.e., when
$P_e(J, J_0)=P_o(J, J_0)$, the low-energy, long-distance behavior of
Eq. (\ref{model}) is universal, and the $XX$ and Heisenberg chains behave
in essentially the same way.

Following Fisher\cite{fisher1}, we introduce a decimation renormalization group
procedure, in which we pick the bond in the system with the largest $J$
(i.e., the strongest bond), say $J_2$ between spins 2 and 3. Since this is
such a strong bond, spins 2 and 3 are likely to form a singlet pair and
become unimportant at low energies (on  scales much smaller than
$J_2$). The major physical effect of the existence of spins 2 and 3 is to
generate an induced coupling between their neighboring spins 1 and 4.
For the $XX$ chain: $\tilde{H}_{1-4}=\tilde{J}_{14}(S_1^xS_4^x+S_1^yS_4^y)$
where $\tilde{J}_{14}=J_1J_3/J_2+O(1/J_2^2)$
and for the Heisenberg chain:
$\tilde{H}_{1-4}=\tilde{J}_{14}\bf{S_1}\cdot\bf{S_4}$
where $\tilde{J}_{14}=J_1J_3/(2J_2)+O(1/J_2^2)$.
The effect
of this decimation procedure is to get rid of the strongest
bond (and also its two neighbors)
in the system, generate a weaker bond between the spins neighboring the
decimated ones, and lower the overall energy scale.
This procedure  becomes asymptotically exact in the low energy
limit\cite{fisher1}.
The new energy cutoff is then lowered to
$\Omega={\rm max}\{\tilde{J}\}$.
Following Fisher and anticipating that the bond distribution
will become broad on logarithmic scales at low energy\cite{fisher1},
we transform to logarithmic variables and
define
$\Gamma=-\log(\Omega/J_0)$
and
$\zeta=\log(\Omega/\tilde{J})/\Gamma$,
so that
both $\Gamma$ and $\zeta$ are positive and a larger $\Gamma$
and a larger
$\zeta$
correspond to a lower energy scale and a weaker bond respectively.
The recursion relations
now become (keeping the leading term only)

\begin{equation}
\tilde{\zeta}_{1-4}=\zeta_1+\zeta_3-\zeta_2+\kappa=\zeta_1+\zeta_3+\kappa
\end{equation}
where we used the fact that $\zeta_2=0$ since $J_2=\Omega$
 is the strongest bond in the system. $\kappa=0$ for the spin-one-half
$XX$ chain and $\log(2)$ for
the spin-one-half Heisenberg chain.
The flow equations for the bond strength
distribution functions $\rho_e(\zeta, \Omega)$ and $\rho_o(\zeta, \Omega)$
in terms of $\zeta$ are then
\begin{eqnarray}
&&{\partial \rho_{e,o}(\zeta, \Gamma)\over
\partial \Gamma}={\partial \rho_{e,o}\over \partial \zeta}
+[\rho_{e,o}(0, \Gamma)-\rho_{o,e}(0, \Gamma)]\rho_{e,o}\nonumber\\
&&+
\rho_{o,e}(0, \Gamma)\int\int{d\zeta_1d\zeta_2}\rho_{e,o}(\zeta_1, \Gamma)
\rho_{e,o}(\zeta_2, \Gamma)\delta(\zeta-\zeta_1-\zeta_2-\kappa).
\label{flow1}
\end{eqnarray}
When $\kappa=0$ (i.e., the $XX$ chain case),
these flow equations are identical to those encountered
in the transverse field Ising
model if we identify the even bonds as the bonds between Ising spins
and odd bonds as the transverse fields\cite{fisher2,notefuture}.
In order to find fixed point solutions of the renormalization group (RG) flow,
it is necessary to
rescale variables. Following Fisher\cite{fisher1,fisher2}, we introduce
the rescaled variable $\eta=\zeta/\Gamma$
and the new distribution function
$Q_{e,o}(\eta, \Gamma)=\Gamma
P_{e,o}(J,\Omega)$.
The flow equations for $Q$ are
\begin{eqnarray}
&&\Gamma{\partial Q_{e,o}\over \partial \Gamma}=Q_{e,o}
+(1+\eta){\partial Q_{e,o}\over
\partial\eta}
+[Q_{e,o}(0,\Gamma)-Q_{o,e}(0,\Gamma)]Q_{e,o}\nonumber\\
&&+Q_{o,e}(0,\Gamma)\int\int
{d\eta_1 d\eta_2}Q_{e,o}(\eta_1)Q_{e,o}(\eta_2)\delta(\eta - \eta_1 - \eta_2
- {\kappa\over\Gamma}).
\label{flow2}
\end{eqnarray}
As Fisher has shown\cite{fisher2,fisher1}, the flow
equations
(\ref{flow2})
have only one generic fixed point\cite{note1}
\begin{equation}
Q_e=Q_o=Q^*(\eta)=e^{-\eta}\Theta(\eta).
\label{fixedpoint}
\end{equation}
This fixed point distribution corresponds to the random spin-1/2 chain
without dimerization, a model studied extensively before\cite{fisher1}.
Going back to the original variable $\zeta$, we find the fixed point
distribution corresponds to
\begin{equation}
\rho(\zeta)={1\over \Gamma}e^{-\zeta/\Gamma},
\end{equation}
i.e., the width of the distribution on the logarithmic scale
grows linearly with the log of the
energy scale $\Gamma$.
For small deviation away from the fixed point $Q_e=Q^*+q_e$ and
$Q_o=Q^*+q_o$,
there is only one {\em relevant} eigenperturbation\cite{fisher2} behaving as
$q_{e,o}(\eta, \Gamma)=q_{e,o}(\eta)\Gamma^\lambda$
with eigenvalue
$\lambda=1$,
and the eigenvector is $q_e=(\eta-1)e^{-\eta}$
and $q_o=-(\eta-1)e^{-\eta}$.
The relevant perturbation, like the fixed point distribution,
is independent of $\kappa$; hence the $XX$ and Heisenberg chains will
behave similarly.  However, non-zero
$\kappa$ does
affect
the irrelevant perturbations.

The relevant perturbation corresponds to the {\em difference} in the
distributions for even and odd bonds.  Therefore we find
that
dimerization is a {\em relevant} perturbation near the RS fixed point, with
eigenvalue $+1$.

For weak dimerization,
the system barely knows that there is a small
difference between even and odd bonds in the early stages of the RG flow.
Both distributions
initially flow toward the RS fixed point solution with a small relevant
perturbation reflecting the extent of dimerization:
\begin{eqnarray}
Q_o(\Gamma)&=&Q^* +\delta\Gamma(\eta - 1)e^{-\eta},\nonumber\\
Q_e(\Gamma)&=&Q^* -\delta\Gamma(\eta - 1)e^{-\eta}
\end{eqnarray}
where $\delta$ characterizes the strength of the dimerization (distance from
criticality).
In general, $\delta$ depends in a complicated way on the shape of
the
original distributions and at what energy scale it is defined.
As the flow away from the RS point continues, the even (odd) bonds get much
weaker
than the odd (even) bonds if originally the even (odd) bonds were only
slightly weaker
than the odd (even) bonds.
The relevant perturbation
grows linearly with $\Gamma$ and becomes of order $O(1)$ as
$\Gamma=\Gamma_0\sim 1/|\delta|$.
The flow equation for the density of spins that have not yet
formed singlets at energy scale $\Gamma$
is
\begin{equation}
{\partial n(\Gamma)\over\partial \Gamma} = -2Q(0,\Gamma)
\label{density}
\end{equation}
Using the RS fixed point distribution
Eq. (\ref{fixedpoint}) in
Eq. (\ref{density}) we find that $n\sim {1/\Gamma^2}$\cite{fisher1,fisher2}
so that when
the relevant perturbation becomes large, the density of active spins is
$n\sim\delta^2$.  The corresponding length scale $L$, which is the typical
distance between the
remaining spins, is
$L_0\sim\Gamma_0^2\sim 1/\delta^2$.
At this stage the existence of dimerization becomes dominant
and under RG
most of the bonds decimated are odd bonds.

The fact that a small difference in the bond distributions
grows as one lowers the energy is physically easy to see. Assume the
odd bonds are slightly stronger than the even bonds in general. Then in the
decimation procedure, it is slightly more likely that
an odd bond gets decimated.
When that happens, typically two intermediate strength neighboring even bonds
also disappear, and a {\em much weaker even bond} gets generated. Hence,
the width of the even bond
distribution
grows faster than
the odd bond distribution,
and its over all strength also decreases faster.
Thus, in the low energy limit, the system can be viewed as a trivially soluble
collection of
uncoupled spin pairs (isolated odd bonds).
We refer to this phase as the Random Dimer (RD) phase.

After renormalization the distribution of odd bonds takes the form
\begin{equation}
\rho_o(\zeta)\sim {1\over \Gamma_0}e^{-\zeta/\Gamma_0}\Theta(\zeta).
\end{equation}
In terms of the original variables the odd bond
distribution is
\begin{equation}
P_o(J)={\alpha\over \Omega_0}({J\over \Omega_0})^{-1+\alpha}
\Theta(1-{J\over \Omega_0}),
\label{distribution}
\end{equation}
where
$\Omega_0=J_0\exp(-\Gamma_0)$
and $\alpha\propto\delta$.

This effective independent pair Hamiltonian
with a power law bond distribution is identical to that introduced by
Clark and Tippie\cite{ct} to explain the low temperature thermodynamics
of the random spin chains. Here we have {\em derived} it using RG from
a realistic model.
The leading temperature dependence
of thermodynamic properties can be determined
by assuming that all spins connected
by bonds with energy greater than the temperature have paired up into singlets
and all spins connected by bonds with energy less than the temperature are
essentially free.  This is a good approximation for broad bond distributions.
In this way, the specific heat and susceptibility in the low temperature
limit can be
easily calculated.
As the temperature
goes to zero, the
the spin susceptibility (in any direction, with possible direction dependent
prefactors) diverges like
$\chi\sim T^{\alpha-1}$,
and the specific heat goes to zero like
$C_v\sim T^{\alpha}$.
The averaged
spin-spin correlation function is short ranged, with the
correlation length (distance between spins)
$\xi\sim |\delta|^{-\nu}\sim |\delta|^{-2}$.
The existence of a divergent magnetic susceptibility
away from the critical point is characteristic of a Griffiths phase.
The divergent susceptibility arises from
magnetically
active gapless excitations.\cite{grifnote}
For the RS phase discussed by Fisher\cite{fisher1},
the averaged
spin-spin correlation function decays as $1/R^2$ at long
distance so the system is critical and one expects a divergent susceptibility.
The
Griffiths phase in the random dimerized spin-one-half chain
is exactly analogous to the
Griffiths phase that appears in the random transverse
field Ising
chain\cite{fisher2}.
If the initial dimerization is large,  the flow begins far from the RS phase
and the bond distribution of the stronger bonds does not flow to a power
law and the gap does not close up.
When this is the case, thermodynamic properties will
depend strongly
on the initial distribution and the susceptibility will remain finite.

The dimer
phase has
a novel kind of topological order that measures the dimerization
of the chain.  The
``string-topological correlation function" is
\begin{equation}
T_{ij}=\left\langle\Psi_0\left|S^z_i
\exp\left[i\pi\sum_{i<k<j}{S^z_k}\right] S^z_j\right|\Psi_0\right\rangle,
\end{equation}
where $|\Psi_0\rangle$ is the ground state.
$T_{ij}$
is similar to the topological correlation function for the spin-one
chain\cite{rn} and it maps onto the spin-spin correlation function of
the transverse field Ising chain.\cite{others}
For a completely dimerized ground state
$T_{ij}=-1/4$ if $i$ is a left spin of a dimer and $j$ is
the right spin of a (possibly different) dimer.
This is because every spin between $i$ and $j$ in the completely dimerized
model is paired up with another
spin between $i$ and $j$.
$T_{ij}=0$ otherwise.
Therefore this special
topological
correlation function is long-ranged although there is only short range
spin-spin correlation, a situation similar to the special kind of
off-diagonal
long range order (ODLRO)
in the fractional quantum Hall effect (FQHE)\cite{gm}.
In the pure case this topological order is closely related to the existence
of a gap\cite{rn}, again just like in the FQHE.

For a general
spin-one-half chain with randomness, we introduce a topological
order parameter
\begin{equation}
T=\lim_{j-i\rightarrow\infty}(\overline{T_{2i,2j-1}}-\overline{T_{2i+1,2j}}),
\end{equation}
where the overbar stands for average over randomness. In the absence of
dimerization, $T$ vanishes. For a random system, $T$ measures the probability
that the two end spins survive decimation until the dimerization becomes large,
and the low energy physics becomes that of the completely dimerized
chain.
This probability is just the square of the density of spins
at the dimerization crossover scale.
Therefore, for small $\delta$, $T$ scales like
\begin{equation}
T\sim -|\delta|^{2\beta}{\rm sgn}(\delta),
\end{equation}
with
$\beta=\nu=2$,
and ${\rm sgn}(\delta)$ is $+$ if even bonds are stronger and $-$ if odd
bonds are stronger. In the absence of randomness $\beta$
is found to be $1/12$\cite{hida}.

The dimerization described in this paper is premeditated because
its magnitude and sign
is determined by the details of the bond distributions.
If the even bonds are stronger than the odd bonds then
it costs less energy to form singlet pairs
over even bonds than over odd bonds.
In the pure
case, premeditated dimerization leads to a unique ground state.
In contrast,
spontaneously dimerized spin chains can dimerize over
even or odd bonds
with equal energy cost.  In the pure case, spontaneous dimerization leads to
degenerate ground states.  The fact that there is a right way and a wrong
way to dimerize a premeditated chain and no wrong way to dimerize a spontaneous
chain leads to profound differences in the response of these systems
to random perturbations.

For the case of premeditated spin chains, the random perturbation
and
quantum and statistical fluctuations will generate regions where
the chain has dimerized in the wrong way.  Between regions of even and
odd dimerization
there has to be a topological soliton which is
essentially a spin unpaired to its neighbors. In the presence of
dimerization, the energy cost of the wrong region
is proportional to
its length (at least at long enough length scales);
 hence the topological solitons are confined by a linear
confining potential and the topological order persists\cite{rs}.
The confinement length in the weak dimerization
limit is essentially the length scale at which dimerization becomes significant
under RG, which is also the spin-spin correlation length.
We hence find that although the gap vanishes in the presence of strong
randomness, the dimer phase is stable and
the topological order persists due to the confinement of topological
defects.
The situation for the FQHE is
similar: incompressibility is lost upon introducing randomness, while
ODLRO survives\cite{sondhi}.   Another example of this
phenomena occurs
in dirty superconductors.  Moderate disorder
can remove the energy gap without
destroying the condensate.

In contrast, topological solitons are unconfined in spontaneously dimerized
systems because there are no wrong regions.  Therefore, unlike the case of
premeditated order, weak
randomness will destroy spontaneously generated
dimerization, despite of the existence of a gap in the absence of
randomness.\cite{future}

We have shown that the dimerized spin-one-half chain is stable
against disorder.  This can be seen as due to the confinement of topological
defects. When the dimerization is weak or randomness is strong
the system is in a
Griffiths phase in which the spin-spin correlation function is short ranged
yet the susceptibility diverges. The
susceptibility and specific heat follow simple power laws with nonuniversal
exponents at low temperatures. The string-topological order is not be
destroyed by disorder. We conjecture that these results apply
to other systems with premeditated (not spontaneously generated) topological
order
such as
the
spin-one chain in the Haldane phase.
We also conjecture that spontaneously generated topological order is unstable
against disorder.
A detailed analysis using a real
space decimation procedure that is appropriate for the spin-1 and
higher spin chains will be presented elsewhere\cite{future}.
Experimentally there has been some work on effects of hole doping in the
spin-1 chains\cite{ditusa}. The effects of hole doping is
probably very different from random bonds in both spin-1 chains and
dimerized spin-1/2 chains because it introduces unconfined topological
defects into the system. Further investigation is underway\cite{future}.

We are extremely grateful to Daniel Fisher and Duncan Haldane for numerous
stimulating discussions and important suggestions. This work was supported
by NSF grants DMR-9224077 and DMR-9416906.


\begin{references}
\bibitem{notexx} Here and throughout the paper
we follow the terminology of Ref. \onlinecite{fisher1}.
The $XX$ chain is called the $XY$ chain in many other contexts.
\bibitem{notefuture} Further insight can be obtained by mapping the $XXZ$
system to a pair of transverse field Ising spin chains. See Refs.
\onlinecite{fisher1,others}.
\bibitem{note1} In the absence of dimerization this is a stable fixed
point that controls all the low energy physics. There are other singular
fixed points that cannot be reached unless the initial distribution is
singular.
See Ref. \onlinecite{fisher1}.
\bibitem{grifnote} The area of the phase diagram where the Griffiths phase
occurs is not really a separate phase.
The gap closes up but there is no level crossing.
\end{references}
\end{document}